\documentstyle [12pt,twoside]{article}
\begin{document}
\small
\vskip .1in
\centerline{\large\bf N-body simulations of galaxies and groups of galaxies}
\centerline{\large\bf with the Marseille GRAPE systems}
\vskip .2in
\renewcommand{\thefootnote}{\alph{footnote}}
\centerline{E. ATHANASSOULA}
\vskip .1in
\centerline{\it Observatoire de Marseille}
\vskip .02in
\centerline{\it 2, place le Verrier}
\vskip .02in
\centerline{\it 13248 Marseille cedex 04, France}
\vskip .02in
\centerline{\it email: lia@paxi,cnrs-mrs.fr, lia@obmara.cnrs-mrs.fr}
\par\noindent
\vskip .15in
\noindent{\small
I review the Marseille GRAPE systems and the N-body
simulations done with them. I first describe briefly the available
hardware and software, their possibilities and their limitations.
I then describe work done on interacting galaxies and groups of
galaxies. This includes simulations of the formation of ring galaxies,
simulations of bar destruction by massive compact satellites, of merging
in compact groups and of the formation of brightest members in
clusters of galaxies.
}
\vskip .15in
\centerline{\bf GRAPE hardware at Marseille Observatory}
\vskip .15in

The idea behind GRAPE systems is at the same time very simple and very 
efficient. It stems from the realisation that most of the CPU time in N-body 
simulations is spent calculating the forces, with only a small percentage 
devoted to the remaining parts, like moving the particles. Thus the group 
around D. Sugimoto and J. Makino realised  
GRAPE (from GRAvity piPE), a card which performs the force calculation
on custom-made chips and which can be put in a 
standard workstation, allowing one to achieve at relatively low cost an 
excellent
CPU performance. A series of such GRAPE boards have been built by the group in 
Tokyo University, starting with GRAPE-1 and evolving steadily to GRAPE-4,
while new members of this family, like GRAPE-5 and GRAPE-6, are in progress. 
For a brief history
of this project and descriptions of the various GRAPE systems see Makino
\& Taiji [1] and references therein. Boards with even numbers 
have high accuracy arithmetic and can be used for collisional simulations, 
where close encounters play an important role in the evolution of the system,
as for globular clusters and planetesimals. Boards with odd numbers have 
lower precision arithmetic and can only be used for simulating 
collisionless systems. 

Two main GRAPE systems are presently working in Marseille Observatory. A 
5-board GRAPE-3AF system, coupled via an Sbus/VMEbus converter to an Ultra 
2/200 front end, and a GRAPE-4 system coupled via a PCI interface to an 
Alpha 500/500 workstation. Since the latter system was only made operational 
a few months before this conference, most of this talk will be devoted to the 
GRAPE-3AF system and the results obtained with it.

Our GRAPE-3AF system has 40 chips in total and gives us a peak speed 
equivalent of more than 20 Gflops. The boards are hardware limited to 
131~072 particles, but it is possible to use them for a much larger 
number by splitting the particles into groups of 131~072 particles 
or less, presenting the groups successively to the board and then 
adding the forces from all the groups on the front end machine.

Since GRAPE-3 boards are meant only for collisionless simulations they use 
low accuracy arithmetic (14 bits for the masses, 20 bits for the positions 
and 56 bits for the forces). As discussed by Athanassoula et al [2], 
this accuracy is sufficient for collisionless simulations.

Doing on-line analysis of the simulation on the same processor as that
used to pilot the GRAPE 
boards would slow down the simulations in an unacceptable way. Thus a 
second processor is necessary. Such on-line analysis can include calculation 
of pattern speeds, amplitudes and shapes of different structures, energy and 
angular momentum exchange between different components etc. The 
processor piloting the boards spawns a task starting the analysis scripts 
at regular intervals. Several tasks, such as calculating the mass still 
bound to a given galaxy, can be carried out much faster on a GRAPE board. In 
such cases the analysis scripts are executed on a smaller GRAPE-3 system,
piloted by another workstation.

A good description of the GRAPE-4 boards and of their performances has been 
given by Makino et al [3]. A description more specific to the Marseille 
system will be given in a forthcoming paper. Since the front end of our 
GRAPE-4 system has no second processor the on-line analysis is spawned
to another 
workstation, acting as a slave to the Alpha 500/500 driving the GRAPE-4 
system.

\vskip .2in
\centerline{\bf GRAPE software at Marseille Observatory}
\vskip .15in

Two codes are routinely used on the Marseille GRAPE-3 systems : a
direct summation
code and a tree code [2, 4]. The latter follows the 
vectorisation scheme proposed by 
Barnes [5]. Thus the particles are first divided into blocks and then 
the tree traversal is executed for a block of particles at a time, rather 
than for each particle separately, as in the standard tree code. The 
optimum number of particles per block depends of course on the number 
of boards, the power of the front end and, to a lesser extent, on the 
number of particles and the tolerance parameter. We find that, for our 
5-board GRAPE-3 configuration and the type of simulations we run,  
7~000 to 15~000 particles per block are a good choice. The CPU time  
necessary for one time step 
increases roughly linearly with the number of particles N. It also increases 
with decreasing tolerance (or opening angle), but the dependence is
less strong than for the 
standard tree code, being particularly small for values of the tolerance 
larger than 0.8. Because of the increased role of the direct summation in 
the force calculation, this tree code is much more accurate than the 
standard one.

The accuracy of the force calculations by GRAPE-3 was tested in [2] with
the help of the MISE/MASE formalism introduced by Merritt [6]. It was 
found that the forces are calculated as accurately as when full precision 
is used on the front end. The reason is that the errors on GRAPE-3 are due 
to round-off and thus can be considered as random. They thus cancel out 
when the force contributions of a large number of particles are added. 
MISE/MASE tests also showed that the accuracy of the treecode is comparable 
to that of the direct summation. That can be easily understood since, in 
the version of the tree code in operation on our GRAPE systems, the 
force from nearby particles is calculated by direct 
summation. 

Further tests include energy conservation during the simulations and 
the comparison of results of runs with 
different number of particles. Finally a few simulations were performed both 
on GRAPE-3 and GRAPE-4 and the comparison shows very good
agreement. Thus it can 
be concluded that GRAPE-3 is well suited for N-body simulations of 
collisionless systems, both because of its accuracy and because of its 
high speed.

It is at present possible to execute three codes on our GRAPE-4 system, a
direct summation and a tree code, both with a constant time step, and a 
direct N-body code with a variable time step. The latter uses an
implementation of the Ahmad-Cohen scheme based on on a fourth order
Hermite integrator [7]. 
A description of their accuracy and performance, as well as a
comparison with those of GRAPE-3, will be given elsewhere.

\vskip .2in
\centerline{\bf Main research areas}
\vskip .15in

Our GRAPE-3 and GRAPE-4 systems are used for N-body simulations in many 
different areas of astronomical research, ranging from dynamics and 
evolution of clusters of galaxies, to the dynamical evolution of 
planetesimals. Most 
of it, however, centers around galaxies and groups of galaxies. Some of 
the latest results are briefly discussed below. To this list should be 
added the study of cusps (in collaboration with Ch. Siopis and H. Kandrup), 
the study of the effect of black holes in the central parts of elliptical 
galaxies (in collaboration with F. Leeuwin), the study of the dynamical 
evolution of planetesimals (in collaboration with P. Barge) and that of 
the gas flows in bars (in collaboration with I. Berentzen and C. Heller).

\vskip .2in
\centerline{\bf Ring galaxies}
\vskip .15in

When a small compact galaxy hits the disc of a target disc galaxy head-on 
and near-vertically, then one or more expanding rings can be generated
[8, 9, 10, 11, 12]. Indeed as the companion approaches the 
target it exerts an extra inwards gravitational force on the disc particles, 
which causes their orbits to contract. This is followed by a rebound,
which, because of the decrease of the epicyclic frequency
with radius, will 
result in a crowding of the orbits and the generation of a high 
amplitude, transient density wave, propagating outwards. We performed a 
number of fully self-consistent N-body simulations of such encounters, both 
on barred and on non-barred target disc galaxies [13]. One or
two transient and 
short-lived rings form, the second considerably after the first one. 
The expansion velocity of the first ring is bigger than that of the
second, and both decrease with time.  
The amplitude, width, lifetime and expansion velocity of the first ring 
are considerably higher for impacts of large mass companions, than for lower 
mass ones. After
the second ring has formed several simulations showed spokes in the
region between the two rings. They are trailing, nearly straight and
short-lived. An example is shown in Fig. 2. Rings formed from such 
head-on encounters
need not be mistaken with those observed at the resonances of disc galaxies. 
Indeed, even when they are symmetric and have no spokes, they have 
considerable expansion 
velocities, which should be detectable spectroscopically.

The Cartwheel is probably the best studied ring galaxy [14, 15, 16,
17, 18, 19]. It has 
two clear rings and several spokes in the region between them. Three small
galaxies can be found in its neighbourhood and one of them should be 
responsible for its structure. Although Higdon [15] proposed that it
is the one farthest from the Cartwheel that is the culprit, a
careful comparison of simulations and observations (Bosma,
Puerari \& Athanassoula, in preparation), taking into account both
the morphological and the kinematical data, argues that it is G2 (in
Higdon's notation) which is to blame.

\vskip .2in
\centerline{\bf Is it possible to destroy a bar without destroying the 
disc it resides in? }
\vskip .15in

An interesting question that can be asked in this context is whether it
is possible for a companion to destroy a bar in a disc galaxy, while not 
destroying the disc. To answer it I first tried trajectories where the
companion, initially on a rectilinear orbit, hit the central part of the 
disc either vertically, or at a skew angle [20]. Such trajectories can bring 
substantial changes to the pattern speed of the bar, as well as to 
the amplitude 
of its m=2 component. The lowering of the m=2 component is in many cases 
very spectacular, so one could easily talk of a bar dissolution. In all these 
cases, however, the disc thickens unnaturally much. I was unable to find 
a case where the disc stayed thin and at the same time the bar was destroyed, 
although I must admit that my search was not exhaustive. 

I then tried a different kind of trajectory [21]. Now the companion 
starts off in a quasi-circular orbit outside the halo, and spirals, via 
dynamical friction, to the central part of the galaxy. If it is 
sufficiently massive and compact, it will lose only a small fraction
of its mass by the time it has reached the center. It will then contribute 
to the bulge population, or, if there was no bulge present before 
the companion fell in, it will form it. Thus the target galaxy will 
evolve along the Hubble sequence, from a late to an early type disc galaxy.
While the companion spirals through the target disc it heats it up and
makes it thicker (cf. [22, 23, 24, 25]). 
On the other hand the target also expands, because the system 
has to conserve angular momentum, so, comparing its axial ratio before
and after the merging, we see that the disc 
has been somewhat, but not much, thickened. The m=2 amplitude of the disc 
decreases very abruptly when the companion reaches the center. At that time 
it increases considerably the central concentration of the galaxy, and, by so 
doing, increases the fraction of irregular orbits present in the disc to the 
detriment of the $x_1$ stable periodic orbits and the regular orbits 
trapped around them [26, 27, 28, 29]. Thus the bar, deprived of its
most ardent 
supporters, will be destroyed. The final stage of such a simulation is 
shown in the two upper panels 
of Fig. 3, the particles initially in the target disc shown in 
blue and the particles initially in the companion in red. Fig. 4
shows separately the particles initially in the target disc (upper two
panels) and the satellite (lower two panels). Note that the satellite,
which was initially spherical, has, after the merging, become oblate.

The situation is totally different in the case of a small mass,  
fluffy companion. In this case the companion looses a lot of its 
mass while spiraling through the disc and no substantial fraction of it will
reach the center. Thus the bar will not be destroyed. Material stripped 
from the 
companion will form a thick disc, thicker than that of the original 
target. The final stage of such a simulation is shown in the lower two panels 
of Fig. 3. Again the particles initially in the target disc are shown in 
blue and the particles initially in the companion in red. Fig. 5
shows separately the particles initially in the target disc (upper two
panels) and the satellite (lower two panels). Note that a substantial
fraction of the companion mass is concentrated along the ends of the bar.

In both the above examples the plane of the companion's orbit 
is the same as that 
of the target's disc. Let us now consider cases where the two planes
are initially at
an angle. Then during the simulation the plane of the target disc 
will tilt [21, 30], so that
the angular momentum of the system be conserved. In the case shown in
[21], and other unpublished simulations, the mass of the companion is
equal to that of the target disc, and the angle by which the disc
tilts is comparable to, although smaller than, the angle between the
orbit of the companion and the plane of the target disc at the
beginning of the simulation. 
The remaining results are as in the case where the companion orbits
initially in the plane of the target's disc, except for an increased 
thickening of both 
the target disc and the disc made by the material shredded from the companion.

\vskip .2in
\centerline{\bf Merging rates in compact groups}
\vskip .15in

Compact groups are groups of a few galaxies, close together in the sky, and 
far from other galaxies or groups of galaxies. Hickson, using a precise 
definition along these lines, catalogued 100 such groups from the Palomar 
sky survey [31]. He has also given a review of the relevant observational data 
about such groups  [32]. One of the main questions that they 
raise is that of their life time. Indeed, if one simply calculates the 
crossing time in such groups from their size and their velocity
dispersion, one finds very low values, from which it has often been
inferred that mergings should occur quite frequently. Thus the question
of why so many such groups are still observed is raised. One possibility is to 
generate such groups continuously from loose groups [33, 34], or add 
new galaxies 
from infalling new material [35]. The first of these alternatives
introduces the problem of the merger remnants, both because the
integrated luminosity of a compact group is three to four times
higher than that of an average
isolated elliptical galaxy, and because the number of such remnants
could be quite high. For the second alternative one or more of the
galaxies in each compact group would have to be the result of previous
mergings, also introducing problems concerning the fraction of
ellipticals in compact groups, as well as their total luminosity. 
Together with Makino \& Bosma [36] I have followed 
another alternative, namely we studied what parameters of the
group affect the merging rate and thus obtained clues about how to form
long-lived compact groups.

Our simulations start with five identical spherical galaxies disposed in 
a compact group which has, in all cases, the same mass and binding energy. 
Two different types of halos have been considered: Either halos around 
individual galaxies (hereafter individual halos, or IH), or halos 
encompassing the whole group and centered at its center (hereafter common 
halos, or CH). We have also considered different luminous-to-dark mass 
ratios, individual halos of different spatial extents, as well as 
different density 
distributions and different kinematics, both for the common halos and the 
distributions of the centers of the galaxies. Once these parameters were 
fixed, we made five different realisations, with different random number 
seeds, in order to allow some, albeit small, statistics, and be able to 
make averages over different realisations. We thus ran over 200 simulations, 
but even so, we are far from covering all possible cases. For all simulations 
we counted the number of galaxies still present in the group as a function 
of time and thus were able to measure merging rates. 

The first question we addressed is whether groups with individual
halos merge faster or slower than 
corresponding groups with common halos, since two contradictory 
results had 
been previously reported in the literature [37, 38, 39]. Indeed, there are two 
different effects influencing the outcome in an opposite sense. On the
one hand dynamical friction is more important in the case of denser
halos, and this should lead to shorter merging times. On the other
hand more massive halos would entail less massive galaxies (since the
total mass of the group is the same in all simulations) and 
therefore less mutual
attractions between them, which would lead to longer merging
times. Which of the two effects is dominant depends on the
configurations. Thus groups with individual halos merge faster than
groups with common halos if the configuration is centrally
concentrated. For less centrally concentrated groups the merging is
initially faster for IH cases, and slower after part of the group has
merged. 

In the case of common halos we find that the more massive the common halo, 
the longer it will take the group to merge. This can be easily understood, 
since the mutual attractions between galaxies is smaller for relatively more 
massive common halos, and, in the extreme case where the masses of 
the individual 
galaxies was zero, then they could be considered as test particles and 
encounters would happen only accidentally, determined by their trajectories 
and cross sections.

Another factor influencing the merging rate is the central
concentration of the configuration. In particular for common halos and
a high
halo-to-total mass ratio, centrally concentrated groups merge
considerably faster. This can be understood because
of the important dynamical friction that galaxies will feel in the
central regions of such centrally concentrated configurations. As far
as the initial kinematics of the group are concerned, groups with
initially cylindrical rotation merge slower.

Taking into account all the effects enumerated above, it is possible
to construct long-lived compact groups. Thus Athanassoula, Makino \&
Bosma [36] followed the evolution of a group with a common halo, a
high halo-to-total mass ratio and a density distribution with little
central concentration and found that the merging occurred only after a
large number of crossing times, corresponding to a time larger than
a Hubble time. This provides a solution to the longevity problem of
compact groups, and could explain why we observe so many of them.

\vskip .2in
\centerline{\bf Formation of brightest cluster members and cD galaxies}
\vskip .15in

We have performed a number of simulations to follow the dynamical evolution 
of groups of 50 to 100 identical spherical galaxies [40, 41, 
and in preparation]. They can be thought of as simulating groups, 
poor clusters, or 
sub-condensations within rich clusters, provided that the dynamical influence 
of the remaining part of the cluster can be neglected. A large variety of 
initial conditions have been considered. This includes the case of individual 
halos (where each halo is centered around a galaxy), or a common halo 
encompassing the whole group or cluster. We have also considered different 
density distributions of the halo material and of the galaxies in the group 
or cluster, different ratios of halo-to-total mass and different initial 
kinematics.

The standard evolution shows important merging in the central regions and 
the formation of a massive central object. The only way to avoid this is to 
consider initial conditions such that the central parts do not contain
any galaxies.
This is obviously artificial, but has the advantage of stressing
the 
role of the initial seed in the formation of the massive central object. It 
also predicts that initial configurations with low central concentration 
should form the central massive object slower than configurations with high 
initial central concentrations, as we were able to confirm with further 
simulations. Let me also note that it is in good agreement with our results 
on compact groups described above. 

Two mechanisms contribute to the formation of the massive central objects.
One is cannibalism of the small galaxies by the big central object (e.g. 
(42, 43]), and the other is accretion onto 
the central object of material that has been stripped off individual small 
galaxies [44, 45, 46]. Both are present in all 
simulations, but to a varying degree, depending on the initial conditions.

We compared the properties of the central massive object with those of 
observed brighter cluster members and found fairly good agreement, 
concerning the morphology, the surface photometry and the kinematics. 
For example
we find that, as is observed [47, 48], the triaxiality is stronger 
in the outer 
than in the inner parts. Also we found that, in the case of non-spherical 
anisotropic initial conditions, the central massive objects ``remember'' 
the orientation of the initial configuration. This should be linked to the 
fact that the orientations of brighter cluster members are not random, but 
correlate with that of the cluster in which they reside [49, 50, 51,
52, 53, 54, 55, 56].

Schombert [57] did photometry of a large sample of brightest cluster members 
and showed that they fall in three classes. In the first class we find 
objects whose projected light profile follows a $r^{1/4}$ over most 
of the galaxy. 
In the second the projected light distribution falls, in the outer parts, 
somewhat below the $r^{1/4}$ that fits the main body of the galaxy, while 
in the third class it is higher than that of the $r^{1/4}$ law. 
Brighter cluster 
members falling in that third category are called cD galaxies, are rather 
frequent and have a light ``halo''. The term ``halo'' is perhaps rather 
unfortunate, since it could lead to confusion with the dark halos around 
galaxies, and it must be stressed that the ``halo'' of cD galaxies is 
luminous material in the outer part of the galaxy, in excess of the 
$r^{1/4}$ law. 

Assuming that the $M/L$ ratio does not depend on radius, we also
calculated projected density profiles  
of the central massive objects formed in our simulations 
and found that they fall in the same three classes as those outlined by 
Schombert for his observed sample. We have then set out to determine 
which properties in the initial conditions determine in which of the 
three classes a central massive object will fall. Although our results 
for this are still preliminary, they nevertheless allow us to make 
a few tentative 
conclusions. Some of the objects we have found  so far in the 
second class 
originated from rather non-spherical initial conditions. 
More interesting are the objects falling 
in the third class. Our simulations suggest a link between the percentage 
of the mass of the central massive object that came via accretion and the 
excess matter in the outer parts, in the sense that we find in the third 
class objects for which a high fraction of their mass is due to accretion. 
This of course shifts the question to what types of initial conditions 
result in a considerable accretion, a question which we are currently 
investigating. In order to have considerable accretion we have 
to have a considerable amount of material which was stripped from 
the initial galaxies. One way of achieving  this is 
to have a quite centrally concentrated common dark matter halo.  
In such a case the galaxies 
passing near the centre of the group or cluster are torn to pieces, thus 
creating the material necessary for the accretion. An alternative way,
which we have not yet verified by numerical simulations, 
would be to have important interactions between the individual galaxies 
before they merge to form the central massive object. Such interactions 
could tear material off the small galaxies by tidal forces, and this 
material could be at later times accreted by the central object. 
The amount of material thus stripped should depend on the 
initial distribution of matter in the individual galaxies, e.g. on 
whether they are  
disc or elliptical galaxies. Thus more elaborate N-body simulations
are necessary to verify this possibility.

\vskip .2in
\centerline{\bf ACKNOWLEDGMENTS}
\vskip .15in
\par\noindent
I am grateful to Albert Bosma and Jean-Charles Lambert, since without
their help and encouragement 
this work would not have been possible. It is also a pleasure to thank all my 
collaborators in the projects mentioned in this paper, and particularly 
Jun Makino, Carlos Garcia-Gomez, Tony Garijo and Ivanio Puerari. I
would also like to thank Philippe Balard for producing figures 2 to 5.
I also
thank the INSU/CNRS, the University of Aix-Marseille I and the
Institut Gassendi (IGRAP) for funds to develop 
the necessary computing facilities. 
The final draft of this manuscript was written at the Newton Institute for
Mathematical Sciences, whose support I acknowledge gratefully.

\vskip .2in
\centerline{\bf REFERENCES}
\vskip .1in

\par\noindent
1. MAKINO, J. \& M. TAIJI.  1998.
Scientific simulations with special purpose computers: \par The GRAPE.
Wiley (Chichester).
\par\noindent
2. ATHANASSOULA, E., A. BOSMA, J.-C. LAMBERT  \& J. MAKINO.
1998. Perfor-{\par}mance and accuracy of a GRAPE-3 system 
for collisionless N-body simulations. Mon. {\par}Not. R. Astron. Soc. 
{\bf 293:} 369-380.
\par\noindent 
3. MAKINO, J., M. TAIJI, T. EBISUZAKI \& D. SUGIMOTO. 1997. GRAPE-4: \par
A Massively Parallel Special-Purpose Computer for Collisional N-body 
Simulations. {\par}Astrophys. J. {\bf 480:} 432-446.
\par\noindent
4. MAKINO, J. 1991. 
Treecode with a Special-Purpose Processor. Publ. Astron. Soc. \par Japan
{\bf 43:} 621-638. 
\par\noindent
5. BARNES, J. 1990. A modified tree code: Don't Laugh; It
Runs. J. Comp. Phys. {\bf 87:} \par 161-170.
\par\noindent
6. MERRITT, D. 1996. Optimal smoothing for N-body codes. Astron. J.
{\bf 111:} 2462-2464.
\par\noindent
7. MAKINO, J. \& S. AARSETH. 1992.
On a Hermite Integrator with Ahmad-Cohen \par Scheme for Gravitational
Many-Body Problems. Publ. Astron. Soc. Japan {\bf 44:} 141-{\par}151.
\par\noindent
8. LYNDS, R. \& A. TOOMRE. 1976. 
On the interpretation of ring galaxies: The binary \par ring system II Hz
4. Astrophys. J. {\bf 209:} 382-388
\par\noindent
9. THEYS, J.C. \& E.A. SPIEGEL. 1976.  
Ring galaxies I. Astrophys. J. {\bf 208:} 650-661.
\par\noindent
10. THEYS, J.C. \& E.A. SPIEGEL. 1977. 
Ring galaxies II. Astrophys. J. {\bf 212:} 616-633.
\par\noindent
11. TOOMRE, A. 1978.
Interacting systems. in {\it The large scale structure of the
Universe}. \par Eds. M.S. Longair \& J. Einasto, I.A.U. Symp. 
{\bf 79:} 109-116. 
\par\noindent
12. APPLETON, P.N. \& C. STRUCK-MARCELL. 1996. 
Collisional ring galaxies. Funda-{\par}mentals of Cosmic Phys. 
{\bf 16:} 111-220.
\par\noindent
13. ATHANASSOULA, E., I. PUERARI \& A. BOSMA. 1997.
Formation of rings by infall \par of a small companion galaxy.
Mon. Not. R. Astron. Soc. {\bf 286:} 284-302.
\par\noindent 
14. HIGDON, J. 1995.
Wheels of Fire I. Massive Star Formation in the Cartwheel Ring \par
Galaxy. Astrophys. J.  {\bf 455:} 524-535.
\par\noindent
15. HIGDON, J. 1996.
Wheels of Fire II. Neutral Hydrogen in the Cartwheel Ring Galaxy.
\par Astron. J. {\bf 467:} 241-260.
\par\noindent
16. AMRAM, P., C. MENDES DE OLIVEIRA, J. BOULESTEIX, \&
C. BALKOWSKI. \par 1998.
The H$\alpha$ Kinematics of the Cartwheel Galaxy. Astron. Astrophys. 
{\bf 330:} 881-893.
\par\noindent
17. STRUCK, C., P.N. APPLETON, K.D. BORNE, \& R.A. LUCAS.  1996.
Hubble \par Space Telescope Imaging of Dust Lanes and Cometary Structures
in the Inner Disk of \par the Cartwheel Ring Galaxy.
Astron. J. {\bf 112:} 1868-1876.
\par\noindent
18. HERNQUIST, L. \& M.L. WEIL. 1993.
Spokes in ring galaxies. Mon. Not. R. Astron. \par Soc. {\bf 261:} 804-818.
\par\noindent
19. STRUCK-MARCEL, C. \& J. HIGDON. 1993.
Hydrodynamic Models of the Cartwheel \par 
Ring Galaxy. Astrophys. J. {\bf 411:} 108-124.
\par\noindent
20. ATHANASSOULA, E. 1996.
Evolution of bars in isolated and interacting disc galaxies 
    \par in {\it Barred Galaxies},
     eds. R. Buta, D. A. Crocker and B. G. Elmegreen. 
     Astron. Soc. \par Pac. Conference series {\bf 91:} 309-321.
\par\noindent 
21. ATHANASSOULA, E. 1996.
The fate of barred galaxies in interacting and merging \par systems; in 
{\it Barred Galaxies and Circumnuclear Activity. Nobel Symposium 
    No. 89},  \par eds. Aa. Sandqvist, P.O. Lindblad, Lecture Notes 
    in Physics, Springer Verlag (Berlin), \par Vol. {\bf 474:} 59-66.
\par\noindent 
22. QUINN, P.J. \& J. GOODMAN. 1986.
Sinking satellites of Spiral Systems. Astrophys. \par J. {\bf 309:} 472-495.
\par\noindent
23. T\'OTH, G. \& J.P. OSTRIKER. 1992. 
Galactic disks, infall
and the global value of $\Omega$. \par Astrophys. J. {\bf 389:} 5-26.
\par\noindent
24. QUINN, P.J., L. HERNQUIST \& D. FULLAGAR. 1993.
Heating of galactic Disks by \par Mergers. Astrophys. J. {\bf 403:} 74-93.
\par\noindent
25. WALKER, I.R.,  J.C. MIHOS \& L. HERNQUIST. 1996.
Quantifying the fragility of \par galactic discs in minor
mergers. Astrophys. J. {\bf 460:} 121-135.
\par\noindent 
26. HASAN, H. \& C. A. NORMAN. 1990. Chaotic orbits in barred
galaxies with central \par mass concentration. 
Astrophys. J. {\bf 361:} 69-77.
\par\noindent 
27. HASAN, H., D. PFENNIGER \& C. A. NORMAN. 1993. Galactic
bars with central \par mass concentrations. Three-dimensional dynamics.
Astrophys. J. {\bf 409:} 91-109.
\par\noindent 
28. NORMAN, C. A., J.A. SELLWOOD \& H. HASAN. 1996.
Bar dissolution and bulge \par formation: An example of secular dynamical
evolution in galaxies. Astrophys. J. {\bf 462:} \par 114-124.
\par\noindent 
29. FRIEDLI, D. \& W. BENZ. 1993. Secular evolution of isolated 
barred galaxies. I \par Gravitational coupling between stellar bars and
interstellar matter. 
Astron. Astrophys. \par {\bf 268:} 65-85.
\par\noindent
30. HUANG, S. \& R.G. CARLBERG. 1997.
Sinking Satellites and Tilting Disk Galaxies. \par
Astrophys. J. {\bf 480:} 503-523.
\par\noindent
31. HICKSON, P. 1982. Systematic properties of compact groups of galaxies.
Astrophys. \par J. {\bf 255:} 382-391
\par\noindent
32. HICKSON, P. 1997.
Compact Groups of Galaxies. Annu. Rev. Astron. Astrophys.
\par {\bf 35:} 357-388
\par\noindent
33. DIAFERIO, A., M.J. GELLER \& M. RAMELLA. 1994. The formation of
compact \par groups of galaxies. I. Optical properties.
Astron. J. {\bf 107:} 868-879 
\par\noindent
34. DIAFERIO, A., M.J. GELLER \& M. RAMELLA. 1994. The formation of
compact \par groups of galaxies. II. X-Ray properties. Astron. J.  
{\bf 109:} 2293-2304.
\par\noindent
35. GOVERNATO, F., P. TOZZI A. \& CAVALIERE. 1996. Small groups of
galaxies: a \par clue to a critical Universe.
Astrophys. J. {\bf 458:} 18-26.
\par\noindent 
36. ATHANASSOULA, E., J. MAKINO \& A. BOSMA. 1997.
Evolution of compact \par groups of galaxies I. Merging rates.
Mon. Not. R. Astron. Soc. {\bf 286:} 825-838.
\par\noindent
37. BARNES, J.E. 1985. The dynamical state of groups of galaxies.
Mon. Not. R. Astron. \par Soc. {\bf 215:} 517-536.
\par\noindent
38. BODE, P.W., H.N. COHN \& P.M. LUGGER. 1992.
Simulations of Compact Groups \par of Galaxies : The effect of the
Dark Matter Distribution.
Astrophys. J. {\bf 416:} 17-25.
\par\noindent
39. ATHANASSOULA, E. \& J. MAKINO. 1995. Simulations of compact groups
of galaxies \par : some preliminary results.
in {\it Compact Groups of Galaxies}, eds.  O. Richter \& K.\par Borne,
A.S.P. conf. series {\bf 70:} 143-149.
\par\noindent 
40. GARC\'{I}A G\'OMEZ, C., E. ATHANASSOULA \& A. GARIJO. 1996.
Dynamical evolu-{\par}tion of galaxy groups: A comparison of two approaches.
Astron. Astrophys. {\bf 313:} 363-\par376.
\par\noindent 
41. GARIJO, A., E. ATHANASSOULA \& C. GARC\'{I}A G\'OMEZ. 1997.
The formation of \par cD galaxies,
Astron. Astrophys. {\bf 327:} 930-946.
\par\noindent 
42. OSTRIKER, J.P. \& S.D. TREMAINE. 1975.
Another evolutionary correction to the \par luminosity of giant galaxies.
Astrophys. J. {\bf 202:} L113-L117.
\par\noindent
43. OSTRIKER, J.P. \& M.A. HAUSMAN. 1977.
Cannibalism among the galaxies - Dynam-{\par}ically produced
evolution of cluster luminosity functions.
Astrophys. J. {\bf 217:} L125-{\par}L129.
\par\noindent
44. GALLAGHER, J.S. \& J.P. OSTRIKER. 1972.
A note on mass loss during collisions \par between galaxies
and the formation of giant systems. 
Astron. J. {\bf 77:} 288-291.
\par\noindent
45. RICHSTONE, D.O. 1975.
Collisions of galaxies in dense clusters. I. Dynamics of
\par collisions of two galaxies.
Astrophys. J. {\bf 200:} 535-547.
\par\noindent
46. RICHSTONE, D.O. 1976.
Collisions of galaxies in dense clusters. II. Dynamical
\par evolution of cluster galaxies.
Astrophys. J. {\bf 204:} 642-648.
\par\noindent
47. PORTER, A.C., D.P. SCHNEIDER \& J.C. HOESSEL. 1991.
CCD observations of \par Abell clusters. V - Isophotometry.
Astron. J. {\bf 101:} 1561-1594.
\par\noindent
48. MACKIE, G., N. VISVANATHAN \& D. CARTER. 1990.
The stellar content of central \par dominant galaxies. I - CCD
surface photometry.
Astrophys. J. Suppl. {\bf 73:} 637-660.
\par\noindent
49. SASTRY, G.N. 1968.
Clusters associated with supergiant galaxies.
Publ. Astron. Soc. \par Pac. {\bf 80:} 252-262.
\par\noindent
50. ROOD, H.J. \& G.N. SASTRY. 1972.
Static properties of galaxies in the cluster Abell \par 2199.
Astron. J. {\bf 77:} 451-458.
\par\noindent
51. AUSTIN, T.B. \& J.V. PEACH. 1974.
Studies of rich clusters II. The structure and \par luminosity function
of the cluster A1413.
Mon. Not. R. Astron. Soc. {\bf 168:} 591-602.
\par\noindent
52. CARTER, D. \& N. METCALFE. 1980.
The morphology of clusters of galaxies.
Mon. \par Not. R. Astron. Soc. {\bf 191:} 325-337.
\par\noindent
53. BINGELLI, B. 1982.
The shape and orientation of clusters of galaxies.
Astron. Astro-{\par}phys. {\bf 107:} 338-349.
\par\noindent
54. STRUBLE, M.F. \& P.J.E. PEEBLES. 1985.
A new application of Binggeli's test for \par large-scale alignment of
clusters of galaxies.
Astron. J. {\bf 90:} 582-589.
\par\noindent
55. RHEE, G. \& P. KATGERT. 1987.
A study of the elongation of Abell clusters I. A \par sample of 37 clusters
studied earlier by Binggeli and Struble \& Peebles.
Astron. Astro-{\par}phys. {\bf 183:} 217-227.
\par\noindent
56. LAMBAS, D.G., E.J. GROTH \& P.J.E. PEEBLES. 1988.
Alignments of brightest \par cluster galaxies with large-scale structures.
Astron. J. {\bf 95:} 996-998.
\par\noindent
57. SCHOMBERT, J.M. 1986.
The structure of brightest cluster members I. Surface \par photometry.
Astrophys. J. Sup. {\bf 60:} 603-693.

\vfill\eject
\centerline{\bf FIGURE CAPTIONS}
\vskip .15in
\par\noindent
FIGS. 1 Schematical representation of the two main GRAPE systems in
Marseille observatory. To the left GRAPE-3 and to the right GRAPE-4.
\vskip .15in
\par\noindent
FIGS. 2 Snapshot from an N-body simulation of the formation of a ring 
galaxy. Only particles initially in the target disc are plotted. Both
rings, as well as the spokes in the region between them, are clearly
visible.
\vskip .1in
\par\noindent
FIGS. 3 One of the final instants from an N-body simulation of a
target disc galaxy
and a satellite, after the merging has been completed. The particles 
of the target disc are shown in blue
and those of the satellite in red, while particles in the halo of the
target are not plotted. Face-on views are given by the left
panels and edge-on views by the right panels. In the simulation shown
in the top two panels the companion was initially massive and
compact. After merging it has lost little of its mass and forms a
bulge in the center of the target disc. In the simulation shown
in the bottom two panels the companion was initially fluffy and less
massive and has been shredded to pieces while spiraling in the
target's disc.
\vskip .1in
\par\noindent
FIGS. 4 Same simulation as the top two panels of Figure 3, but now the
particles from the target disc and companion are shown separately, the
target disc in the top two panels and the companion in the two bottom
ones.
\vskip .1in
\par\noindent
FIGS. 5 Same simulation as the bottom two panels of Figure 3, but now the
particles from the target disc and companion are shown separately, the
target disc in the top two panels and the companion in the two bottom
ones.

\vfill\eject\end{document}